\documentclass[aps,prl,superscriptaddress]{revtex4}
\usepackage{amssymb}

\usepackage{amsmath}
\usepackage{graphicx}

\begin{document}

\title{Traversal Times for Random Walks on Small-World Networks}
\author{Paul E. Parris}
\affiliation{Consortium of the Americas for Interdisciplinary Science and Department of
Physics and Astronomy, University of New Mexico, Albuquerque, NM 87131}
\affiliation{Department of Physics, University of Missouri-Rolla, Rolla, MO 65409}
\author{V. M. Kenkre}
\affiliation{Consortium of the Americas for Interdisciplinary Science and Department of
Physics and Astronomy, University of New Mexico, Albuquerque, NM 87131}
\date{\today }

\begin{abstract}
We study the mean traversal time $\tau $ for a class of random walks on
Newman-Watts small-world networks, in which steps around the edge of the
network occur with a transition rate $F$ that is different from the rate $f$
for steps across small-world connections. When $f\gg F,$ the mean time $\tau 
$ to traverse the network exhibits a transition associated with percolation
of the random graph (i.e., small-world) part of the network, and a collapse
of the data onto a universal curve. This transition was not observed in
earlier studies in which equal transition rates were assumed for all allowed
steps. We develop a simple self-consistent effective medium theory and show
that it gives a quantitatively correct description of the traversal time in
all parameter regimes except the immediate neighborhood of the transition,
as is characteristic of most effective medium theories.
\end{abstract}

\maketitle

\section{Introduction}\label{intro}

There has been intense recent interest in statistical and dynamical
properties associated with random networks possessing so-called ``small
world'' (SW) properties, i.e., networks whose bonds possess a large degree
of local clustering, but with a relatively small minimal path length
connecting nodes of the system. At least some of this interest has arisen
from the realization that algorithmically constructed small-world networks
(SWNs) introduced by Watts and Strogatz \cite{ws}, and by Newman and Watts %
\cite{nw}, appear to share many statistical features with social networks,
and thus form a useful topological substrate on which to model dynamical
processes relevant to the medical and social sciences \cite{kuperman}. Power
grids and information networks may also possess properties associated with
small-world or scale-free networks\cite{barabasi}, and even certain aspects
of polymers may be understood in terms of an underlying small-world
structure \cite{rm}.

Independently of their possible applications to a wide range of problems in
the social, physical, and information sciences, small-world networks and
their extensions are inherently interesting because they provide a class of
disordered structures similar to those that have been studied in the
condensed matter literature for many years \cite%
{simulation,kirkpatrick,odagakilax,hauskehr,parris}. There is an extensive
literature, for example, in which the dynamics of random walkers diffusing
through various kinds of disordered media have been studied through a
variety of methods, including numerical simulations \cite{simulation},
percolation ideas \cite{kirkpatrick}, effective medium theories (EMTs), \cite%
{kirkpatrick,odagakilax,hauskehr,parris} and others.

In the present paper we study traversal times associated with random walks
on the Newman-Watts small-world networks (NW-SWN's). These structures are
defined on a ring of $N$ sites, each of which has bonds that connect it to
its $2K$ nearest neighbors. On this translationally-invariant lattice a
(generally) disordered small-world structure is imposed, by associating with
probability $q$ a SW connection along each of the remaining $N(N-2K-1)/2$
bonds in the system. For a given value of $q$, there are on average $%
n_{sw}=q(N-2K-1)/2\sim qN/2$ small-world bonds per site in the system, a
parameter that allows for a useful comparison between SWNs of different
size. Except for the endpoints at $q=1$ and $q=0$, where the system is
perfectly ordered, the network forms a disordered system on which various
forms of dynamics can be studied.

The new feature of our analysis, which is based upon a master equation for
the site occupation probabilities $P_{n}\left( t\right) $, is that we take
the jump rate $f$ associated with SW connections to be independent of the
jump rate $F$ associated with steps around the edge of the network. Models
of this sort might arise, for example, in attempts to design, modify, or
optimize existing information networks in order to reduce the mean access
time, by incorporating a small number of fast connections into an existing
random network already possessing a large number of other (perhaps slower)
connections. By introducing this simple extension to the situation in which
all allowed steps occur with the same probability (and hence same transition
rate), our theory is able to reveal interesting universal behavior that
occurs for $f\gtrsim F,$ including a percolation transition that occurs at a
critical fraction of SW bonds. Such a transition, which is clearly a
consequence of a well-known result by R\'{e}nyi and Erd\"{o}s \cite{erdos}
for random graphs, was not observed in earlier random walk studies of the
NW-SWN \cite{stroud,blumen}, in which only equal jump probabilities ($f=F$)
were considered. The transition that we observe numerically and analyze
using a simple self-consistent effective medium theory, is critically
dependent on the fact that the topological disorder associated with the
small-world connections is ``quenched''. We show, for example, that an
exactly solvable model with ``annealed disorder'', in which the walker
decides at each step whether to move along the edge or to a randomly chosen
site elsewhere in the network, shows no such transition as a function of the
branching ratio between steps around the edge of the ring and steps
associated with small-world type ``shortcuts''.

The rest of the paper is laid out as follows. In the next section we
introduce a master equation description of continuous time random walks on
Newman-Watts small-world networks, and review the relation between the
Green's functions associated with those equations to the mean time $\tau $
for a random walker on such a network to traverse the system. We then
present numerical calculations of the mean traversal time and show that for $%
f\gtrsim F$ there is a transition in the underlying transport mode, between
motion that is dominated (and limited ) by edge diffusion and motion that
takes place predominantly among small-world connections in the system. In
that section also we consider a model with annealed disorder and show that
it does \emph{not} display the transition exhibited by the model with
quenched disorder. In Sec. \ref{EMT} we analytically 
consider the traversal time
for NW-SWN's using a simple self-consistent effective medium (EMT) theory
similar to that which has been used to understand other disordered systems.
As in previous work, we find that EMT provides an accurate and
computationally efficient means for calculating transport properties as a
function of the underlying parameters of the system, except in the immediate
vicinity of the transition where the underlying transport mechanism changes
abruptly. We are thus able to use it to reliably investigate, in regimes
away from such points, the manner with which various transport properties
scale with system size and connectivity.

\section{Master Equations, Traversal Times, and Percolation}\label{master}

We consider a continuous time random walk with symmetric,
translationally-invariant jump rates $F_{n,m}=F_{n-m}$ governing transitions
between sites $n$ and $m$ on a 1-dimensional ring of $N$ sites. Although it
is not crucial in the analysis that follows, for specificity we assume that $%
F_{n,m}=F$ is constant and non-zero only between a given site and its $2K$
closest neighbors on the ring, which we will refer to collectively as
``neighbor sites''. The small-world structure of the system is imposed by
associating with probability $q$ a nonzero SW hopping rate $f$ along each of
the $N(N-2K-1)/2$ non-neighbor bonds in the system. For $q=1$ and $q=0$, the
system is perfectly ordered; in the former, all sites are connected, in the
latter, connections occur only between sites on the periphery of the ring.
For intermediate values of $q$ a particle moving along the ring encounters
sites from which it can move across the system without taking steps to its $%
2K$ nearest neighbors. The small-world connections thus superimpose a
``random graph'' structure on the otherwise translationally invariant
lattice.

Random walks on any single NW-SWN constructed in this manner can be
described by the master equation 
\begin{equation}
\frac{dP_{m}}{dt}=\sum_{n=-K}^{K}F\left( P_{m+n}-P_{m}\right) +\sum_{n\neq
-K}^{K}f_{mn}\left( P_{m+n}-P_{m}\right)  \label{ME}
\end{equation}%
for the probability $P_{m}(t)$ for the particle to be at site $m$ at time $t$%
, where we assume periodic boundary conditions so that $P_{m+N}=P_{m}$. In (%
\ref{ME}), the first sum on the right describe steps around the edge of the
ring, while the last term describes SW transitions. In this last term, the $%
f_{mn}$ are symmetric random rates that in any given realization equal $f$
with probability $q$ and equal zero with probability $1-q$.

A full solution to the problem involves obtaining the set of propagators, or
Green's functions $g_{m,n}(t)$, the solutions to Eq. (\ref{ME}) for each
initial site $n$ at which the particle may start. These quantities, or their
Laplace transforms $\tilde{g}_{m,n}\left( \varepsilon \right)
=\int_{0}^{\infty }g_{m,n}\left( t\right) e^{-\varepsilon t}dt,$ are readily
computed for moderately sized systems $N\lesssim 10^{3}$ using standard
numerical techniques. For example, rewriting the equations of motion (\ref%
{ME}) in the form%
\begin{equation}
\frac{dP_{m}}{dt}+ \sum_{n}A_{mn}P_{n}\left( t\right) = 0  \label{Adefine}
\end{equation}%
which along with Eq. (\ref{ME}) implicitly defines a transition matrix $%
\mathbf{A}$ with elements $A_{mn}$, we can write the Green's function in the
time domain as the $m$-$n$ element of an exponential 
\begin{equation}
g_{m,n}\left( t\right) =\left[ e^{- \mathbf{A}t}\right] _{m,n}
\end{equation}%
of the matrix $\mathbf{A}.$ Similarly, we can express its Laplace transform
as the corresponding element

\begin{equation}
\tilde{g}_{m,n}\left( \varepsilon \right) =\left[ \left( \varepsilon +%
\mathbf{A}\right) ^{-1}\right] _{m,n}
\label{resolvent}
\end{equation}%
of the resolvent matrix $\left( \varepsilon +\mathbf{A}\right) ^{-1}$, which
is easily computed using matrix inversion routines.

Numerical solutions of this sort can be used to find information about the
evolution of the probabilities $P_{m}\left( t\right) ,$ their moments, or
other quantities that characterize transport in the system. In the condensed
matter literature, e.g., much attention has been placed on calculating the
diffusion constant $D$ which in an infinite Euclidean network characterizes
the asymptotic linear growth of the mean square displacement $\langle
n^{2}\left( t\right) \rangle \sim Dt$ of an ensemble of random walkers.
Insofar as we are interested in the properties of finite networks, in which
the mean-square displacement always saturates at long times, we focus here
on calculating properties that may be more relevant to applications for
which small-world concepts are currently employed. Specifically, we focus
here on calculating the mean traversal time $\tau $, which we define to be
the earliest time, on average, that a random walker visits the point on the
ring the farthest from where it started. For a walker starting at site $n$
at $t=0,$ the mean time $\tau _{m,n}$ to arrive at an arbitrary site $m$ for
the first time is the first moment of the probability density $F_{m,n}(t)$
for a walker to first arrive at site $m$ at time $t$ for these initial
conditions, i.e.,%
\begin{eqnarray}
\tau _{m,n} &=&\int_{0}^{\infty }F_{m,n}\left( t\right)
t\;dt=-\lim_{\varepsilon \rightarrow 0}\frac{d}{d\varepsilon }%
\int_{0}^{\infty }e^{-\varepsilon t}F_{m,n}\left( t\right) \;dt  \label{FPT}
\\
&=&-\lim_{\varepsilon \rightarrow 0}\frac{d\tilde{F}_{m,n}\left( \varepsilon
\right) }{d\varepsilon },
\end{eqnarray}%
where in the second line we have expressed the result in terms of the
Laplace transform $\tilde{F}_{m,n}\left( \varepsilon \right) $ of $%
F_{m,n}\left( t\right) $. The first passage probability density $%
F_{m,n}\left( t\right) $ is related to elements of the Green's functions
described above through the relation%
\begin{equation}
g_{m,n}\left( t\right) =\int_{0}^{t}g_{m,m}\left( t-t^{\prime }\right)
F_{m,n}(t^{\prime })\;dt^{\prime }
\end{equation}%
which physically expresses the probability for the walker to be found at $m$
at the current time $t,$ in terms of its probability to have arrived at that
site for the first time at some earlier moment $t^{\prime }$, and the
probability that it is now at that site given that it did so. From the
convolution theorem for Laplace transforms it then follows that 
\begin{equation}
\tilde{F}_{m,n}\left( \varepsilon \right) =\frac{\tilde{g}_{m,n}\left(
\varepsilon \right) }{\tilde{g}_{m,m}\left( \varepsilon \right) },
\end{equation}%
so that we can write Eq. (\ref{FPT}) as 
\begin{equation}
\tau _{m,n}=-\lim_{\varepsilon \rightarrow 0}\frac{d}{d\varepsilon }\left[ 
\frac{\tilde{g}_{m,n}\left( \varepsilon \right) }{\tilde{g}_{m,m}\left(
\varepsilon \right) }\right] .  \label{tauFPT}
\end{equation}%
\begin{figure}[t]
\centering \includegraphics[height=2.5in]{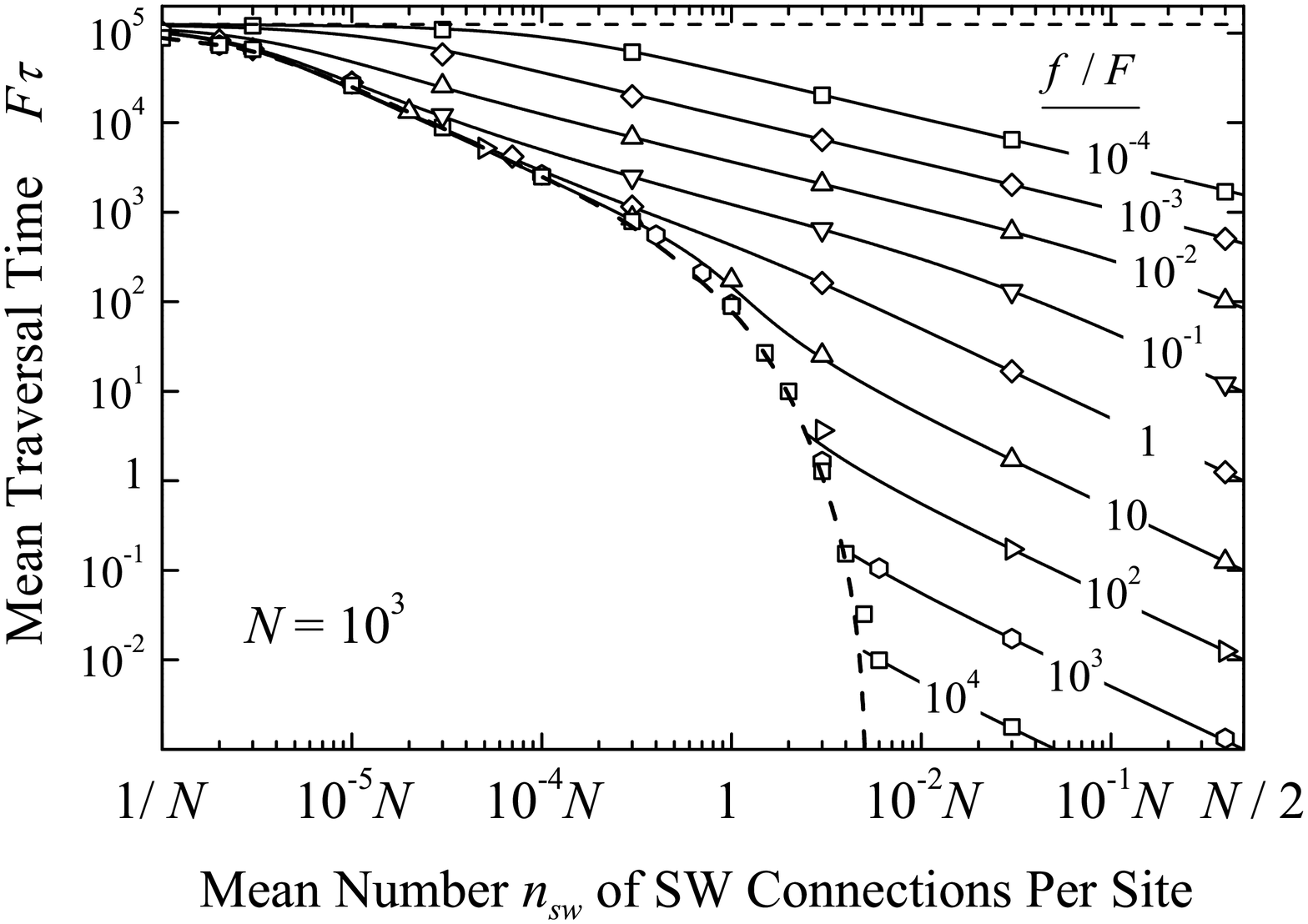}
\caption{Mean traversal time $\protect\tau $ as a function of the mean
number $n_{sw}$ of small-world connections per site on a SWN of $10^{3}$
sites, with values of $f/F$ as indicated.}
\label{fig1}
\end{figure}
The mean first passage time is thus readily computed from a numerical
solution to the Laplace transformed Green's 
function of the system (\ref{resolvent}), or
equivalently, through the resolvent of the matrix $\mathbf{A}$.

In Fig.\ \ref{fig1} we display the results of a numerical calculation of the
mean traversal time 
\begin{equation}
\tau =N^{-1}\sum_{m}\langle \tau _{m+\frac{N}{2},m}\rangle 
\end{equation}%
scaled by the average nearest neighbor hopping time $F^{-1},$ averaged over
an ensemble of 100 $K=1$ NW-SWN's with $N=10^{3}$ as a function of the
average number $n_{sw}$ of small-world bonds per site, for different values
of the ratio $f/F$. As expected, the mean traversal time $\tau $ decreases
monotonically as the number $n_{sw}$ of small-world connections increases.
In the limit in which $n_{sw}<N^{-1},$ i.e., in the absence of any
small-world shortcuts, the mean traversal time reduces to the mean time $%
\tau _{0}$ that it takes for the walker to diffuse around the ring, for
which $2F\tau _{0}\sim \left( N/2\right) ^{2}.$ Also, as we might expect,
for very small values of $f/F,$ only modest decreases in the traversal time
occur with the addition of small world network connections.

For values of $f\geq F,$ however, the situation appears very different,
there generally being a very strong decrease in the traversal time near $%
n_{sw}\sim 1-5$, and a strong collapse of the numerical data onto a single
curve for values of $n_{sw}<1$. In this region SW connections are sparse,
but fast. For $n_{sw}<1$ and $f\gg F,$ the SW connections act as
short-circuits, but transport across the system is still limited by
diffusion within segments of the ring that are free of shortcuts, because
there is not in this regime a percolating path of small-world connections
spanning the system. However, as shown by R\'{e}nyi and Erd\"{o}s \cite%
{erdos}, in a completely random graph a percolating path develops (as $%
N\rightarrow \infty )$ as $n_{sw}$ approaches one from below. Thus, as $%
f/F\rightarrow \infty ,$ and $N\rightarrow \infty ,$ percolation of the
random-graph part of the NW-SWN network leads to a critical change in the
traversal time near $n_{sw}\sim 1$.

It is interesting to note that other transport properties of the NW-SWN's,
including scaling properties of the diagonal element $\tilde{g}_{00}\left(
\varepsilon \right) $ of the average green's function\cite{stroud}, and the
distinct number of sites visited $S\left( t\right) $ by a walker on the
network\cite{stroud} have been studied previously, but only for the case in
which $f=F,$ i.e., in which hops associated with small-world connections
occur with the same transition rate as steps along the periphery. As we see
from the current study, however, the case $F=f$ lies at the edge of a
parameter regime in which the behavior of the system, as a function of the
number of small-world connections, changes drastically.

It is natural to ask whether the quenched disorder occurring in the
small-world networks is an essential element for the observed transport
threshold to occur, or whether a simpler \emph{annealed disorder }model with
uniform trans-network connections could exhibit a similar transport
transition as a function of their strength. We show explicitly below that a
simple annealed disorder model does not exhibit such a transition.
Specifically, we study an exactly soluble model in which a walker at any
site takes steps to its neighbors on the edge of the ring with rate $F,$ as
before, and takes steps to one of $n$ randomly chosen non-neighbor sites on
the ring with rate $f,$ but in such a way that the $n$ randomly chosen sites
to which it may move are not fixed in time, but are selected anew \emph{each
time the particle visits the site.} Thus, in this model, if we take $%
n=2n_{sw}$ the branching ratio between steps associated with short cuts
across the system and those around the edge is, on average, the same as in
the NW-SWN model with quenched disorder already considered. Since the \emph{%
a priori} probability of the particle moving to any other site on the ring
other than its neighbors is the same, the dynamics on this annealed disorder
model is equivalent to one in which the short cuts allow the walker to move
to any one of its non-neighbors with a uniform rate 
\begin{equation}
\omega =2n_{sw}f/N_{n},
\end{equation}%
where $N_{n}=N-2K-1$ is the number of non-neighbor sites to which it can
move. Thus in any given site, the walker moves to one of its neighbors with
probability $2KF/(2KF+2n_{sw}f)$ and to a non-neighbor with probability $%
n_{sw}f/(2KF+2n_{sw}f).$ The resulting master equation for this model, 
\begin{equation}
\frac{dP_{m}}{dt}=\sum_{n=-K}^{K}F\left( P_{m+n}-P_{m}\right) +\sum_{n\neq
-K}^{K}\omega \left( P_{m+n}-P_{m}\right)  \label{AnnealedME}
\end{equation}%
is translationally invariant and readily solved by introducing Fourier
transformed probabilities $P^{k}\left( t\right) =\sum_{n}P_{n}\left(
t\right) e^{ikn}$. We find, e.g., that for this annealed disorder model the
Laplace transformed Green's functions, $\tilde{g}_{m}^{A}\left( \varepsilon
\right) =\tilde{g}_{m,0}^{A}\left( \varepsilon \right) ,$ are given by the
relation%
\begin{equation}
\tilde{g}_{m}^{A}\left( \varepsilon \right) 
=\frac{N\omega }{N\varepsilon \left(
\varepsilon +N\omega \right) }+\tilde{G}_{m}\left( \varepsilon +N\omega
\right)  \label{Anneal g}
\end{equation}%
in which 
\begin{equation}
\tilde{G}_{m}\left( \varepsilon \right) =N^{-1}\sum_{k}\frac{e^{ikm}}{%
\varepsilon +\left( F-\omega \right) A_{k}},  \label{annealG}
\end{equation}%
\begin{equation}
A_{k}=\sum_{n=1}^{K}2(1-\cos kn),
\end{equation}%
and the sum in (\ref{annealG}) is over all wavevectors $k=2\pi n/N$ in the
Brillouin zone.

\begin{figure}[t]
\centering \includegraphics[height=2.5in]{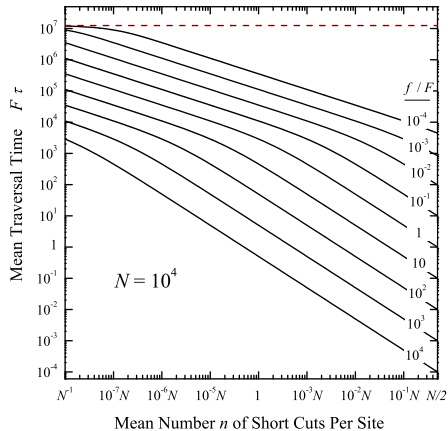}
\caption{Mean traversal time $\protect\tau $ as a function of the mean
number $n$ of trans-network connections per site for the annnealed
disorder model on a system with $10^{4}$ sites, with values of $f/F$ as
indicated, showing the failure of the annealed disorder to capture the
transport transition observed to occur on real small-world networks.}
\label{fig2}
\end{figure}

In Fig.\ \ref{fig2} we show how the mean traversal time $\tau $ scales in the
annealed disorder model as function of $n$ (now considered a continuous
parameter), for various values of $f$ for a system with $N=10^{4}.$ Clearly,
although the limiting diffusive behavior seen in the quenched disorder model
of Fig.\ \ref{fig1} as $n_{sw}\rightarrow 0$ and $f\rightarrow 0$ is 
the same as
that observed in Fig.\ \ref{fig2}, the previously observed behavior for $f\gg F$,
and in particular the transition near $n_{sw}\sim 1$ is completely lacking
in the model with annealed disorder. Thus, as we might expect, the
underlying percolative behavior is clearly a property associated with a
system with quenched disorder. We note in passing that the annealed disorder
model described above can be viewed as an \emph{approximation }to the
quenched disorder model in which the random matrix $\mathbf{A}$ is replaced
by its ensemble average $\langle \mathbf{A}\rangle ,$ i.e., in which%
\begin{equation}
\langle \left( \varepsilon +\mathbf{A}\right) ^{-1}\rangle \approx \left(
\varepsilon +\langle \mathbf{A\rangle }\right) ^{-1}.
\end{equation}%
For disordered systems, approximations of this type are known to give
results that are often qualitatively very bad, particularly in percolative
systems, where they typically wash out any transition that occurs. It is not
surprising, therefore, that the two models have very different behavior. 

It is clear that the annealed disorder model fails to capture the essential
features of the transport transition that occurs in the NW small world
networks. However, the idea of replacing a disordered system with a
translationally-invariant one that captures, on large time and length
scales, the macroscopic transport properties of the ensemble of systems that
it replaces is venerable and has a long history in the condensed matter
literature. Indeed, the goal behind a significant body of theoretical work
on various kinds of disordered systems has been to construct an appropriate,
self-consistently determined effective medium, whose properties captures
qualitative features of the actual system under consideration. In the next
section, we explore this idea, and introduce a simple effective medium
theory that does, in fact, provide an excellent quantitative prediction of
the traversal time for the NW-SWN in all parameter regimes except the
immediate neighborhood of the critical point.

\section{Self-consistent Effective Medium Theory of the Traversal Time}%
\label{EMT}

Theoretical justification for the search for a translationally-invariant
effective medium whose properties capture the essential features of the
ensemble of disordered systems that they replace (and which capture the
large time and length scale properties, typically, for any member of the
ensemble) lies with the fact that \emph{average} transport properties of the
ensemble are, in fact, translationally invariant. For example, the
ensemble-averaged probabilities $p_{m}\left( t\right) =\langle
P_{m}(t)\rangle $ associated with any fixed initial condition evolve in a
translationally invariant way, and therefore obey translationally invariant
equations of motion which, if we knew what they were, would serve to define
the effective medium that we seek. Indeed, the goal of effective medium
theory is to self-consistently determine properties associated with this
average evolution. To this end, we consider the simplest set of linear,
homogenous, and translationally-invariant equations of motion that reflect
the structure of the original equations, and the symmetry properties of the
underlying ensemble. In particular, the Laplace transforms $\tilde{p}%
_{m}\left( \varepsilon \right) $ of the average probabilities $p_{m}\left(
t\right) $ we take to obey the equations%
\begin{equation}
\varepsilon \tilde{p}_{m}\left( \varepsilon \right) -p_{m}\left( 0\right)
=\sum_{n=-K}^{K}F\left( \tilde{p}_{m+n}-\tilde{p}_{m}\right) +\sum_{n\neq
-K}^{K}\tilde{w}\left( \varepsilon \right) \left( \tilde{p}_{m+n}-\tilde{p}%
_{m}\right) .  \label{EMA ME}
\end{equation}%
Here, $\tilde{w}\left( \varepsilon \right) $ is a frequency-dependent rate,
equivalently, a memory function in the Laplace domain \cite{footnote3},
connecting pairs of sites on the network capable of being connected, in any
realization, by a small-world rate $f$. In these effective medium equations
of motion, transport \emph{around} the ring edge is characterized by the
same rate $F$ that obtains throughout the ensemble, but $\tilde{w}\left(
\varepsilon \right) $ must be determined from self-consistent
considerations. Note that these EMT equations of motion have exactly the
same form as the Laplace transform of those describing the annealed disorder
model of the last section, except that in that model the rate $\tilde{w}%
\left( \varepsilon \right) =\omega $ was simply set equal to its ensemble
average, while in the current treatment $\tilde{w}\left( \varepsilon \right) 
$ is to be determined self-consistently through other considerations.

Thus, the theoretical tasks are two: (i) the self-consistent determination
of $\tilde{w}\left( \varepsilon \right) $ as a function of $N$ and $q$ (or $%
n_{sw}$), and (ii) the determination of transport properties arising through
the solutions to (\ref{EMA ME}). Both tasks require calculation of the
effective medium propagators $\tilde{g}_{m}^{e}\left( \varepsilon \right) $,
i.e., the solutions to (\ref{EMA ME}) for a walker initially at the origin, $%
p_{m}\left( 0\right) =\delta _{m,0}$. Since the structure of the equations
of motion (\ref{EMA ME}) are the same as those of the annealed disorder
model, the form of the effective medium Green's function associated with (%
\ref{EMA ME}) is the same as (\ref{Anneal g}), with $\omega $ replaced by $%
\tilde{w}\left( \varepsilon \right) :$%
\begin{equation}
\tilde{g}_{m}^{e}\left( \varepsilon \right) =\frac{N\tilde{w}}{N\varepsilon \left(
\varepsilon +N\tilde{w}\right) }+\tilde{G}_{m}\left( \varepsilon +N\tilde{w}%
\right)  \label{EMA g}
\end{equation}%
where 
\begin{equation}
\tilde{G}_{m}\left( \varepsilon \right) =N^{-1}\sum_{k}\frac{e^{ikm}}{%
\varepsilon +2\left( F-\tilde{w}\right) A_{k}}.  \label{ring G}
\end{equation}

\begin{figure}[t]
\centering \includegraphics[height=2.5in]{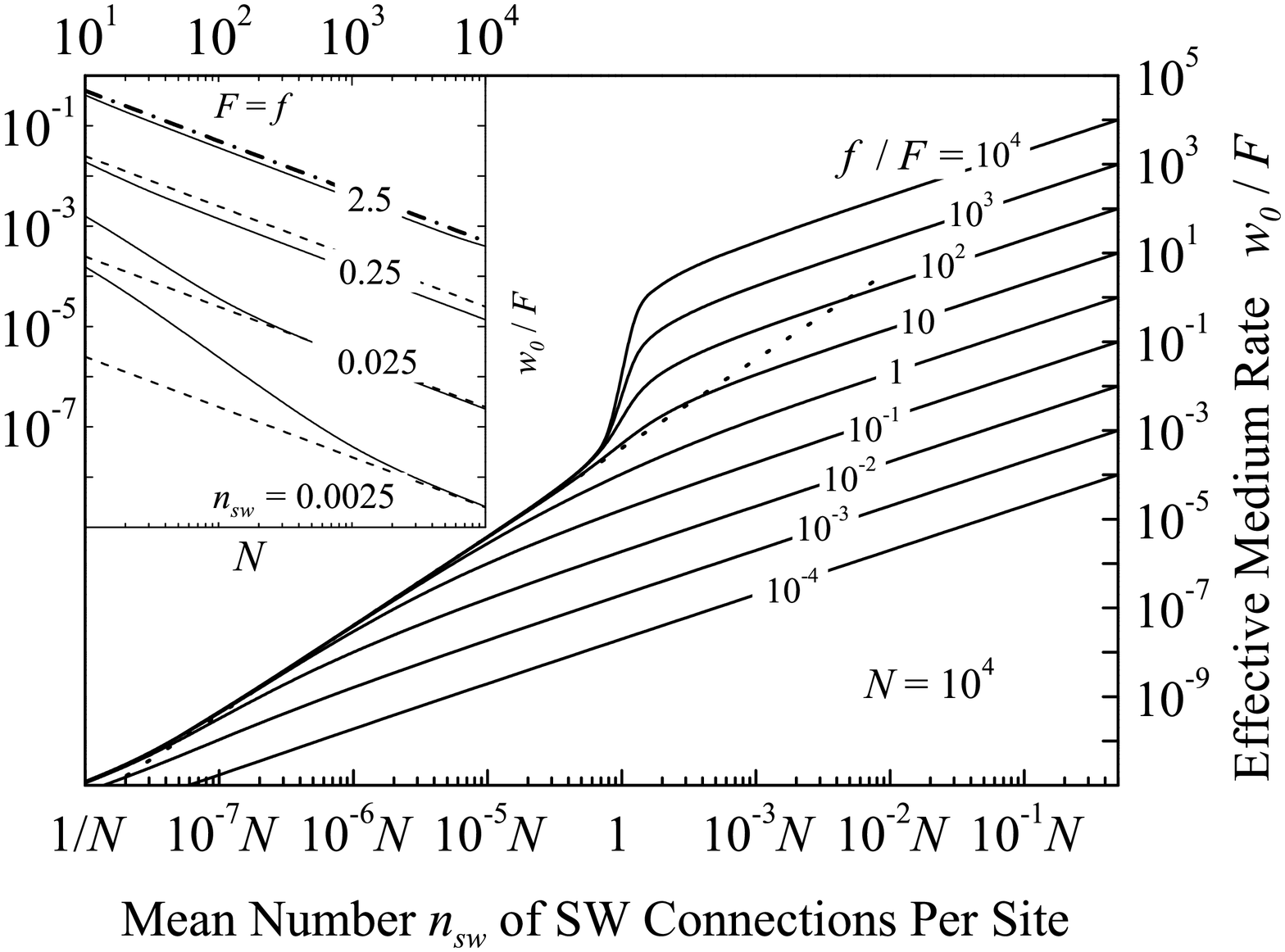}
\caption{Effective medium parameter $w_0$ as a function of the mean number $%
n_{sw}$ of small-world connections per site on a SWN of $10^{4}$ sites, with
values of $f/F$ as indicated. Inset: Effective medium parameter $w_0$ as a
function of the number $N$ of sites in the network, for the case $f = F$.}
\label{fig3}
\end{figure}

Using (\ref{EMA g}) we now develop a self-consistent expression for $\tilde{w%
}\left( \varepsilon \right) $. We proceed as in other EMTs \cite%
{kirkpatrick,odagakilax,hauskehr,parris}, by embedding in the effective
medium a typical ``fluctuation from the average'' associated with the
disorder, and then requiring that upon averaging over the distribution
associated with this fluctuation we recover the effective medium in which it
was embedded. Thus, we consider the bond between the origin and some other
non-neighbor site $n$ on the ring, and replace the effective medium bond $%
\tilde{w}\left( \varepsilon \right) $ between those two sites with an actual
rate $f_{0,n}=f_{n}$ ($=0$ or $f$), drawn from the ensemble. For a particle
placed at the origin, the resulting $\tilde{p}_{m}(\varepsilon )$ obey 
\begin{equation}
\varepsilon \tilde{p}_{m}\left( \varepsilon \right) -\delta
_{m,0}=\sum_{n=-K}^{K}F\left( \tilde{p}_{n+m}-\tilde{p}_{m}\right)
+ \sum_{n\neq -K}^{K}\tilde{w}\left( \varepsilon \right) \left( \tilde{p}%
_{m+n}-\tilde{p}_{m}\right) -\tilde{\Delta}_{m}\left( \varepsilon \right)
\label{defect}
\end{equation}%
where $\tilde{\Delta}_{m}\left( \varepsilon \right) =\left( \delta
_{m,0}-\delta _{m,n}\right) \left( f_{n}-\tilde{w}\right) \left( \tilde{p}%
_{0}-\tilde{p}_{n}\right) $. The solution to the set of equations (\ref%
{defect}) can be written in terms of (\ref{EMA g}) as 
\begin{equation}
\tilde{p}_{0}=\tilde{g}_{0}^{e}-\frac{\tilde{\alpha}_{n}\tilde{\gamma}%
_{n}^{2}}{1+2\tilde{\alpha}_{n}\tilde{\gamma}_{n}},  \label{Defect solution}
\end{equation}%
where $\tilde{\alpha}_{n}\left( \varepsilon \right) =f_{n}-\tilde{w}\left(
\varepsilon \right) ,$ and $\tilde{\gamma}_{n}\left( \varepsilon \right) =%
\tilde{g}_{0}^{e}\left( \varepsilon \right) -\tilde{g}_{n}^{e}\left(
\varepsilon \right) .$

We now impose self-consistency, and require that, upon averaging (\ref%
{Defect solution}) over the dichotomous distribution 
\begin{equation}
\rho \left( f_{n}\right) =\left( 1-q\right) \delta \left( f_{n}\right)
+q\delta \left( f_{n}-f\right)
\end{equation}%
of the rates $f_{n}$, on the one hand, \emph{and} the location $n$ of the
small-world connection, on the other, we recover the propagator for the
effective medium. For this to occur, the second term on the right hand side
of (\ref{Defect solution}) must average out to zero. This leads to the
self-consistent condition%
\begin{equation}
\sum_{n=K+1}^{N/2}\frac{q\left( f-\tilde{w}\right) \tilde{\gamma}_{n}^{2}}{%
1+2\left( f-\tilde{w}\right) \tilde{\gamma}_{n}}=\sum_{n=K+1}^{N/2}\frac{%
\left( 1-q\right) \tilde{w}\tilde{\gamma}_{n}^{2}}{1-2\tilde{w}\tilde{\gamma}%
_{n}},  \label{SCC}
\end{equation}%
from which $\tilde{w}\left( \varepsilon \right) $ can be determined. For $%
q=1 $ and $q=0,$ respectively, the solutions reduce to the exact results. 
For $q\neq 0$ or $1$, Eq.(\ref{SCC}) can be solved
numerically without having to perform large matrix inverses.

\begin{figure}[tbp]
\centering \includegraphics[height=2.5in]{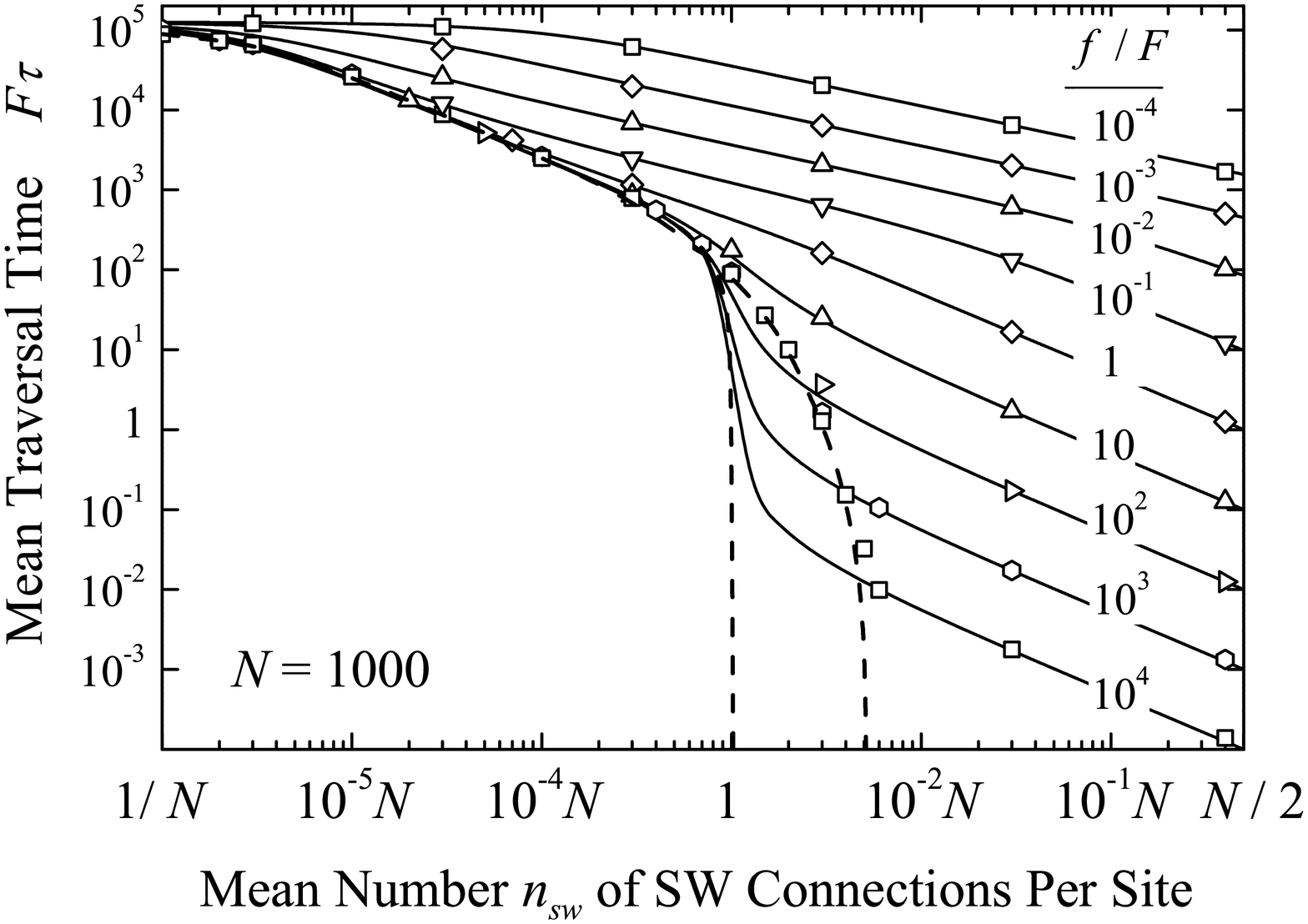}
\caption{Average time to first traverse the network as a function of the
mean number $n_{sw}$ of small world connections per site, on a SWN of $%
10^{3} $ sites, with values of $f/F$ as indicated. Solid lines are from EMT,
data points are the same as in Fig.\ \protect{\ref{fig1}}}
\label{fig4}
\end{figure}

Because we are primarily interested in properties of the network associated
with large length scales, and thus with times much longer than that required
for a single hop, we are interested in the Laplace domain on the behavior of
the system for small values of the 
Laplace variable $\varepsilon $ as in Eq. (\ref{tauFPT}).
Accordingly, we present in Fig.\ \ref{fig3} a plot of the zero-frequency
effective medium parameter $w_{0}=\lim_{\varepsilon \rightarrow 0}\tilde{w}%
\left( \varepsilon \right) $ as a function of the number $n_{sw}=qN/2$ of
small-world connections per site, for different values of the SW hopping
rate $f$, on a NW-SWN with $N=10^{4}$ sites. On the right hand side of the
figure, at sufficiently large values of $n_{sw}$ for any value of $f,$ the
effective medium parameter $w_{0}$ appears well-described by the formula $%
w_{0}\sim qf\sim 2n_{sw}f/N.$ As we move to values of $n_{sw}\sim 1$ and
less, there is a convergence of the curves with $f>F,$ similar to that which
appeared in Fig.\ \ref{fig1} for the traversal time of the NW-SWN, thus giving hope
that the self-consistent treatment introduced here provides a qualitatively
correct description of transport in this disordered system. After some
analysis of the results, we find that in the region $1>n_{sw}$, where the
different curves with $f>F$ converge, the parameter $w_{0}$ is independent
of $f$ and well-characterized by the relation $w_{0}\sim Nq^{2}F\sim
4n_{sw}^{2}F/N$, depicted as a dashed line in the figure. As $N\rightarrow
\infty ,$ the region of the main part of Fig.\ \ref{fig3} associated with
this universal behavior extends downward and to the left, so that for any $f$
there will be a value $n_{sw}(f)$ below which $w_{0}$ will scale as $%
4n_{sw}^{2}F/N$. In the inset of Figure \ref{fig3} we show how this scaling
relation is obeyed as a function of system size for large $N$ and small $%
n_{sw},$ for the case in which $f=F$. Dotted lines in the inset indicate the
relation $w_{0}\sim 4n_{sw}^{2}F/N,$ which the curves with small $n_{sw}$
approach for large $N$. At moderate values of $n_{sw}>1,$ scaling crosses
over to the form $w_{0}\sim 2n_{sw}f/N,$ represented by the dot-dashed line,
shown for $n_{sw}=2.5$. For very small values of $n_{sw}\;\lesssim $ $%
N^{-1}, $ uninteresting deviations from SW scaling occur in the regime where
most members of the ensemble have no SW connections.

This interesting behavior of the effective medium parameter $w_{0}$ leads to
two important questions. Does the behavior seen in the parameter $w_{0}$
lead to a corresponding effect on actual transport observables? And do the
predictions of EMT for those observables describe the actual system of
interest? At least in terms of the mean traversal time the answer to both of
these questions is affirmative. The derivatives and small-$\varepsilon $
limit in (\ref{tauFPT}) can be explicitly taken for the effective medium
propagators (\ref{EMA g}), leading to the simplification%
\begin{equation}
\tau _{m}=N\left[ \tilde{G}_{0}\left( Nw_{0}\right) -\tilde{G}_{m}\left(
Nw_{0}\right) \right] .
\end{equation}

In Fig.\ \ref{fig4}, we plot the mean traversal time $\tau =\tau _{N/2}$ as
a function of $n_{sw}$ for a system with $N=10^{3}$ sites. Solid lines in
this figure are predictions of EMT, while the data points are the same as
those appearing in Fig.\ \ref{fig1}. In Fig.\ \ref{fig4} we see that the
effective medium theory predicts a transition in the vicinity of $n_{sw}=1,$
as well as a collapse of the numerical data for large $f/F$ onto a single
curve for $n_{sw}<1$. For all values of $f$, traversal times saturate for
very small $n_{sw}$ to a value such that $2F\tau \sim \left( N/2\right) ^{2}$%
, consistent with pure diffusion around the edge of the ring. In the region $%
1\gtrsim n_{sw}\gtrsim N^{-1}$ where the data collapse occurs for $f>F,$ the
traversal time of the numerical data and the EFT are both described by the
functional relation $\tau \sim (2qF)^{-1}=N/4n_{sw}F$, which is a reflection
of the mean number of steps $F\tau \sim (2q)^{-1}$ the particle must take
along the edge of the ring before it encounters a SW shortcut across the
system.

\begin{figure}[t]
\centering \includegraphics[width=3in]{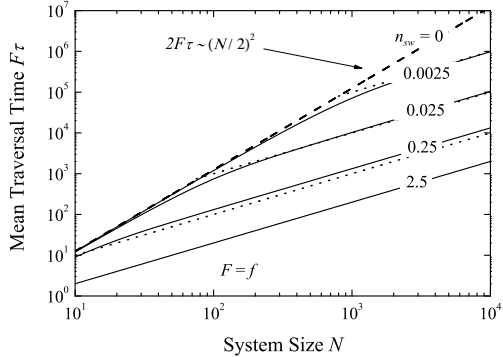}
\caption{Mean traversal time as a function of the number $N$ of sites in the
network, for $f = F$. Dashed line is the limiting case for pure diffusion,
solid lines are the results of EMT. Dotted lines indicate scaling
relationships discussed in the text.}
\label{fig5}
\end{figure}

Comparison of the data and the solid curves in Fig.\ \ref{fig4} shows that,
as in other EMTs, the theory derived above appears to be numerically
accurate for all parameters except those near the underlying percolation
transition, the effect of which only becomes apparent for large values of $%
f/F$ in the neighborhood of $n_{sw}\sim 1$. Provided that we avoid this
particular regime, i.e., the region bounded by the dashed curves in Fig.\ %
\ref{fig4}, we can reliably use our EMT to calculate accurately other
properties of the system. The two dashed curves in that figure represent
limiting cases for very large $f/F>10^{8}$ for the EMT (on the left) and the
numerical results (on the right). The discrepancy between EMT, and the exact
numerical results does not decrease significantly with increasing $N$. Thus,
the EMT of the present paper is typical of those for other disordered
systems, which often capture the essential behavior of the system, but fail
to accurately reproduce critical properties. Of course the main advantage of
the EMT we have derived is that we can now extend its calculations to larger
networks (i.e., larger values of $N$) for which numerical solutions to the
individual equations of motion (\ref{ME}) become prohibitively difficult. As
an example, we present in Fig.\ \ref{fig5} a plot of traversal time $\tau $
as a function of system size $N$. This plot has been computed using EMT for
fixed values of $n_{sw},$ for the case $f=F$, which is accurately described
by EMT for all $n_{sw}$, as is evident from Fig.\ \ref{fig4}. Clearly, the
predicted scaling $\tau \sim N$, shown as dotted lines in Fig.\ \ref{fig5},
obtains for any value of $n_{sw}$ for sufficiently large values of $N$.
Indeed, for small $n_{sw},$ the time to traverse the system follows closely
the diffusive result $\tau \sim N^{2}$ until the system size is sufficiently
large that there is a significant probability to encounter a shortcut before
reaching the other side of the ring via diffusion around the periphery.

\section{Summary}\label{summary}

In this paper we have considered the mean traversal time $\tau $ for a class
of random walks on Newman-Watts small-world networks, in which steps around
the edge of the network occur with a transition rate $F$ that is different
than the rate $f$ for steps across small-world connections. Using numerical
calculations of the Green's function we obtained present numerical data for
the mean traversal time and show that for $f\gtrsim F$ there is a transition
in the underlying transport mode, between diffusion limited and connection
limited motion. We showed that a model with annealed disorder does not
display a transition reflective of the underlying percolation transition
occurring on the random graph part of the network. We then developed a
self-consistent effective medium (EMT) theory for the traversal time that
provides an accurate and computationally efficient means for calculating
this quantity except in the immediate vicinity of the transition. As a
result, we have been able to use EMT to determine how various transport
properties scale with system size and connectivity on the Newman-Watts small
world networks.

This work was supported in part by the National Science Foundation under
Grants DMR-0097204, DMR-0097210, INT-0336343, and DARPA-N00014-03-1-0900.

\end{document}